\documentstyle[epsfig]{aipproc}
\begin{document}
\title{STM Studies of\\Synthetic Peptide Monolayers}

\author{David J. Bergeron$^*$, Wilfried Clauss$^{*{\dag}}$, Denis L. Pilloud$^{\ddag}$,\\
P. Leslie Dutton$^{\ddag}$, and Alan T. Johnson$^*$}
\address{$^*$Department of Physics and Astronomy\\
$^{\ddag}$Johnson Research Foundation, Department of Biochemistry and Biophysics, School of Medicine\\
University of Pennsylvania, Philadelphia PA 19104\\
$^{\dag}$On leave from Institute of Applied Physics,
University of T\"ubingen, 72074 T\"ubingen, Germany\\
}

\maketitle

\begin{abstract}

We have used scanning probe microscopy to investigate self-assembled
monolayers of chemically synthesized peptides.
We find that the peptides form a dense uniform monolayer, above 
which is found a sparse additional layer.
Using scanning tunneling microscopy, submolecular resolution can be obtained, revealing the alpha helices
which constitute the peptide.
The nature of the images is not significantly affected by the incorporation
of redox cofactors (hemes) in the peptides.
\end{abstract}

\section*{Introduction}
Synthetic peptides are an exciting new form of engineerable electronic material.
Chemical synthesis allows peptides to be synthesized with arbitrary amino acid 
sequences.  It is therefore possible to design molecules with minimal complexity,
yet which fold to a desired structure and can incorporate redox cofactors.

We have studied a redox peptide based on the prototype of Robertson {\it et al.} \cite{robertson94}  
with minor modifications \cite{gibney97}.
The peptide consists of a pair of 2-helix dimers, each joined by disulfide
bonds (see Figure 1).  The sulfurs allow the peptide to form a self-assembled monolayer (SAM) \cite{ulman}
when dispersed onto a gold substrate.  Alternatively, a peptide
SAM can be deposited onto a SAM of dimercaptoalkane linker.
The interior of the four-helix bundle contains four ligation
sites for binding metalloporphyrin hemes.   Heme redox
centers found in natural proteins play an important role in
electron transport \cite{moser92,moser96}.  In viewing synthetic peptides as a novel
electronic material, we consider hemes as the basis for engineering
the material's electronic properties. 
Using a scanning probe microscope, we have investigated the morphology of 
peptide SAMs,
both on linker and on bare gold, with various numbers of hemes incorporated into the peptides.

\section*{Experimental Method}
Substrates for self-assembly are made by evaporating
gold onto cleaved mica squares which have been heated for 3 hr to 300 $^{\circ}$C.  
Annealing these in a gas flame produces atomically
flat Au(111) terraces with typical sizes around 100 nm.

When a linker layer is used, the substrates are immersed in a
solution of dimercaptoalkanes, which form a self-assembled
monolayer on the gold surface.  The samples are rinsed in isopropanol, and
then immersed overnight in a 200 $\mu$M solution of
peptide.  Alternatively, the peptide self assembly step is performed directly on
the bare annealed gold.
are 
Our primary tool for studying the peptide SAMs is an Omicron Beetle 
Scanning Tunneling Microscope (STM).
In this work, all
images were taken under ambient conditions, using
tunneling impedances ranging from 1 to 15 G$\Omega$.
The Omicron Beetle can also be operated as an Atomic Force Microscope (AFM).
AFM is not sensitive to the conductivity of the sample and 
forces exerted by the tip can be substantially lower than in STM.

\section*{Results}
As seen in Figures 2 and 3, our STM data show a uniform monolayer of peptides.
The separation between nearest neighbors is typically 4-5 nm, while the observed vertical
corrugation of the layer is only a few \AA ngstroms. 
Occasional depressions
in the layer suggest voids in the monolayer, but are more likely
etch pits in the gold substrates.  Such etching is 
characteristic of thiol self-assembly processes \cite{poirier97}.

Our images (see Figure 2) show that there is a sparse second layer of
peptides which are probably not covalently bound into the monolayer, and
which are mobile under the influence of the STM imaging.  In the STM
images, this layer appears as horizontal streaks up to 5 nm in length
which persist for one or more scan lines of the image.  We infer that
a peptide molecule is present in a location long enough to be imaged for a 
few scan
lines, and then is removed by the tip to another location.  
By imaging
with AFM, we can reduce the forces disturbing molecules, and we
find that the second layer remains immobile on the surface.  
Figure 2 shows the manifestations of this layer in both STM and AFM images.

We have studied peptide SAMs using low `conventional' tunneling
impedances of 1-10 G$\Omega$, as well as relatively high values of
13-15 G$\Omega$.  By imaging at higher impedances, we raise
the tip further above the sample surface.  The effect of this 
on our images is to reveal the 4-helix substructure of the peptides.
The $\alpha$ helices appear as round features separated by 2-3 nm
from their nearest neighbors (Figure 3).  
Images taken with various current setpoints and bias voltages suggest that the
enhanced resolution is due to raising the impedance (and therefore
the tip-sample separation), and not due to the increase in voltage alone.
We conclude that at low impedances, the bottom of the STM tip is actually
passing within the peptide layer, whereas at larger impedances, it is
raised to near the surface of the monolayer.  

When a linker layer is present between the peptide monolayer and the gold
substrate, we find that imaging at low tunneling impedances leads to etching
of the gold substrate.  This is probably a result of linker molecules
which bind to the Pt/Ir STM tip, and then pull gold atoms out of the
the substrate.  At higher tunneling impedances (above 13 G$\Omega$), the
tip is further above the linker layer, and this
process is not observed.  In this case, the images are essentially identical to
those obtained without a linker present.

Finally, we have investigated peptide SAMs with the number of hemes per peptide
varying from 0 to 4.  Furthermore, a monolayer was made with a mixture of both
peptides containing hemes and those without.  Until now, we have
not observed any effect in the images or tunneling spectra which could
be attributed to the presence of hemes.
Observation of such an effect will
be the subject of future research.

This work is supported by Penn's Laboratory for Research on the Structure
of Matter, and NSF MRSEC IRG \#DMR-96-32598. W.C. thanks the Deutsche Forschungsgemeinschaft. 
P.L.D. receives funding from NIH grant \#GM-41048.
A.T.J. is supported by a David and Lucile Packard Foundation Fellowship.

\begin{figure}[H] 
\caption{(a) Schematic of the four-helix peptide, shown with
4 hemes ligated.
(b) The iron(III) protoporphyrin heme molecule.}
\label{fig1}
\end{figure}
\begin{figure} 
\caption{Left image: 100x100nm STM image of a peptide SAM.
The Au(111) terrace structure dominates the image. 
The grains on the largest visible terrace are peptides in the SAM.
The monolayer is punctuated by dark spots and bright
horizontal streaks.  The spots are attributed to etch pits in the gold layer,
while the streaks are caused by a mobile second layer of peptides. 
V=1.1 V, I=220 pA. 
Right image:  1 $\micron$ x 1 $\micron$ AFM image of another  
peptide SAM showing sparse second layer of peptides.  The image is
created from the feedback signal (phase), which yields an image with a `shaded'
appearance.  
The peptides appear enlarged due to finite tip size. Some of the underlying
gold grain structure is also visible.}
\end{figure}
\begin{figure}[H] 
\caption{Images of peptide SAMs at low (left) and high (right) tunneling impedances,
showing increased resolution at higher tip-sample distance.  Features in the left
images are full peptides.  In the right image, individual $\alpha$ helices are observed.
Most streaks due to the mobile layer of peptides have been removed in both images.
Left: V=1.5 V, I=200pA. Right: V=2.3V, I=120pA.
Both images are 43nm x 43nm with a vertical range of 0.6nm.}
\end{figure}

\end{document}